\documentclass{article}
\usepackage{spconf,amsmath,graphicx}
\usepackage{amsfonts}
\usepackage{multirow}
\usepackage{hyperref}
\usepackage{subfigure}
\usepackage{booktabs,setspace}
\usepackage{threeparttable}
\ninept
\usepackage{setspace}

\usepackage{amssymb,CJK, indentfirst,balance,appendix}

\usepackage{gensymb}
\usepackage{enumitem}
\usepackage{color}
\usepackage{bbding}
\usepackage{algorithm,algpseudocode}
\title{THE CUHK-TENCENT SPEAKER DIARIZATION SYSTEM FOR THE ICASSP 2022 MULTI-CHANNEL MULTI-PARTY MEETING TRANSCRIPTION CHALLENGE}
\name{\begin{tabular}{c}Naijun Zheng$^1$, Na Li$^2$, Xixin Wu$^1$, Lingwei Meng$^1$, Jiawen Kang$^1$, Haibin Wu$^3$ \\
Chao Weng$^2$, Dan Su$^2$, Helen Meng$^1$\end{tabular}}

\address{$^1$The Chinese University of Hong Kong \\
        $^2$Tencent AI Lab \\
        $^3$National Taiwan University
        }

\begin{document}
\ninept
\maketitle
\begin{abstract}
This paper describes our speaker diarization system submitted to the Multi-channel Multi-party Meeting Transcription (M2MeT) challenge,
where Mandarin meeting data were recorded in multi-channel format for diarization and automatic speech recognition (ASR) tasks.
In these meeting scenarios, the uncertainty of the speaker number and the high ratio of overlapped speech present great challenges
for diarization.
%Meanwhile, the limited size of the real training dataset also blocks the development of the very deep neural networks.
%We attempt to utilize the spatial information to enable better speaker identification capability.
%By estimating the speakers' direction of arrival(DOA), our systems can better detect the activity of the target speakers in presence of overlapped speech. 
Based on the assumption that there is valuable complementary information between acoustic features, spatial-related and speaker-related features, we propose a multi-level feature fusion mechanism based target-speaker voice activity detection (FFM-TS-VAD) system to improve the performance of the conventional TS-VAD system. 
Furthermore, we propose a data augmentation method during training to improve the system robustness when the angular difference between two speakers is relatively small.
%We also investigate a limiting case for our system, where the angular difference between two speakers is relatively small.
%In this case, only using spatial features becomes less effective than using the conventional acoustic speaker embeddings like i-vectors.
%To achieve mutual complementary, we applied feature fusion in our systems.
%Furthermore, we propose a mask-based data augmentation method during training to make the network more robust when using spatial features.
We provide comparisons for different sub-systems we used in M2MeT challenge. 
Our submission is a fusion of several sub-systems and ranks second in the diarization task.
% experimental results on the corpus show the effectiveness of our proposed FFM-TS-VAD system.

\end{abstract}
\section{Introduction}
\label{sec:intro}

Speaker diarization is the task of determining ``who spoke when" given a long audio \cite{anguera2012speaker}. 
It is a crucial component for meeting transcription which aims to solve the problem of ``who speaks what at when" in real-world multi-speaker conversations. 
Recently, speaker diarization for meeting scenarios has attracted increasing research enthusiasm. However, the lack of large, public real meeting datasets has been a major obstacle for advancing diarization technologies. 
To address this problem, the M2MeT challenge \cite{yu2021m2met} offers a sizeable corpus for research in speaker diarization and multi-speaker ASR. 

This paper describes the CUHK-TENCENT diarization system submitted to Track 1 of the M2MeT challenge. 
Since overlapped speech spreads throughout the multiple speaker conversations \cite{boakye2008overlapped}, we focused our efforts on how to handle overlapped speech in the diarization task. 
We employed a VBx system with an overlapped speech detector (OSD) as our baseline. 
To identify speakers more accurately for overlapped speech, we modified the conventional TS-VAD system \cite{Medennikov2020} by replacing the speaker detection layer with self-attention mechanism to better predict target speaker's activity.
% Specifically, according to the initial diarization annotations obtained from a VBx baseline system, conventional TS-VAD systems \cite{Medennikov2020} were employed to improve the performance on overlapped speech.
Since the spatial information associated with speakers' locations can also be used for distinguishing speakers, we estimated the direction of arrival (DOAs) of target speakers and employed our proposed target-DOA system \cite{njzheng2021target} to detect their activities. % given short-term signals. 
Based on the assumption that there is valuable complementary information between acoustic features, spatial-related and speaker-related features, we propose a multi-level feature fusion mechanism based target-speaker voice activity detection (FFM-TS-VAD) system to improve the performance of the conventional TS-VAD system and our previous target-DOA system.
Furthermore, we propose a data augmentation strategy during training to improve the system robustness when the angular difference between two speakers is relatively small.
Experimental results on evaluation set demonstrate the superiority of FFM-TS-VAD system compared to other systems.     

The rest of the paper is organized as follows:
Section \ref{sec:dataset} introduces the M2Met challenge.
Section \ref{sec:overview} presents the overview of our submitted systems.
Section \ref{sec:ffm} describes the proposed FFM-TS-VAD system with feature fusion and the data augmentation strategy.
The experimental setup is given in section \ref{sec:experiments}.
The results are discussed in section \ref{sec:results}.
Finally, conclusions are drawn in section \ref{sec:con}.

\section{M2MeT challenge Dataset}
\label{sec:dataset}

The M2MeT challenge provides a sizeable real-recorded Mandarin meeting corpus called AliMeeting for speaker diarization task (Track1) and ASR task (Track2).
The corpus provides far-field data recorded by 8-element microphone arrays and near-field data collected by headset microphones respectively, where only far-field data is provided for testing.
% \begin{itemize}
% \item Training dataset: 212 sessions (104.75 hours), 456 speakers.
% \item Evaluation dataset: 8 sessions (4 hours), 25 speakers.
% \item Test dataset: 20 sessions (10 hours).
% \end{itemize}
The AliMeeting corpus contains 240 meetings in total, where 212 meetings (104.75 hours) for training, 8 meetings (4 hours) for evaluation and 20 meetings (10 hours) for testing.
Large number of speakers, various meeting venues and high overlap ratios present huge challenges in real meeting scenarios.
The average speech overlap ratios of training and evaluation sets reach 42.27\% and 34.76\% respectively.
For far-field recording, the microphone-speaker distance ranges from 0.3 to 5.0 meters.
For each session, the number of speakers ranges from 2 to 4 and the speakers are required to stay in the same positions during recording.

There are two sub-tracks in the speaker diarization task.
In sub-track 1, the system is required to be built from fixed constrained data, i.e., AliMeeting, AISHELL-4 \cite{fu2021aishell} and CN-Celeb \cite{fan2020cn} corpus.
In sub-track 2, extra data is allowed for system training.
More details about the challenge data and evaluation metrics can be found in \cite{yu2021m2met}.

\section{System overview}
\label{sec:overview}
For this challenge, we submitted a system that fuses of four types of sub-systems. A brief description of each type of sub-system is shown as follows:

\subsection{Baseline system}
We used a conventional pipeline system composed of the VBx clustering modules \cite{landini2021analysis} and an overlapped speech detector as our baseline system.
Specifically, the BeamformIt \cite{anguera2007acoustic} technique was applied to covert multi-channel signals into single-channel signals.
Based on the (estimated/oracle) VAD bounds, the voiced segments were sliced into sub-segments with a fixed-length window of 1.5 seconds and a sliding overlap of 1.25 seconds.
Then a speaker embedding extractor based on an 101-layer Res2Net \cite{gao2019res2net} was used to extract embeddings using the inputs of 40-dimensional logarithmic Mel FBank features.
Given the embeddings, the spectrum clustering (SpC) method followed by the VBx algorithm \cite{diez2019analysis} was employed to  perform clustering.
A post-processing step called re-clustering was applied to further refining the number of speakers by combining the very similar clusters according to their cosine distances \cite{landini2021analysis}.

Since the conventional pipeline system cannot handle overlapped speech, a deep network composed of SincNet convolution layers and LSTM layers \cite{Bredin2021} was trained for VAD and overlapped speech detection (OSD) simultaneously.
The heuristic method \cite{landini2021analysis} was used to handle the overlapped speech by finding a different but closest speaker along the time axis as the second speaker.
% The output of our baseline system will be used as initial diarization annotations for the following system to extract the accurate embeddings and DOA of speakers.

\subsection{Modified TS-VAD system}
\label{ssec:tsvad}

The TS-VAD system proposed by Rao et al. \cite{Medennikov2020} used the i-vectors as the anchors to identify the speakers within overlapped speech, where the i-vectors were estimated based on the initial diarization annotations from the baseline system.
In the original network, bi-directional LSTM (BLSTM) layers were used to learn the temporal relationship between the i-vectors and the acoustic features.
Considering the strong capability of self-attention mechanism to capture the global speaker characteristics in diarization systems \cite{fujita2019end, lin2020self}, we added an encoder layer after the convolutional layers  to obtain the acoustic feature embeddings.
The following two BLSTM layers used for speaker detection were replaced with another two encoder layers.
The encoder layers are the same as the transformer encoders \cite{vaswani2017attention} but without positional encoding.
Two i-vector extractors were trained using 64-dimensional logarithmic Mel FBank features and 40-dimensional MFCC features, respectively.

\subsection{Target-DOA system}
\label{ssec:tdoa}

Since the direction of speakers' sound can also be used for discriminating speakers, we proposed a Target-DOA system using the spatial features as inputs.
Firstly, the DOAs of target speakers were estimated according to the initial diarization annotations.
Then, the angle features (AF) \cite{chen2018multi, yu2020end} were computed based on the DOAs and multi-channel signals, the corresponding computation process is given in Section \ref{ssec:doa}. 
Finally, the magnitude spectrum and AF were concatenated as the input to the network consisting of temporal convolutional network (TCN) blocks \cite{luo2019conv}.
% , and $N$ output ports give the activity information of $N$ target speakers given their estimated DOA, which are denoted as ${\bf{\hat{y}}^\text{vad}}=\{\hat{y}^\text{vad}_i\}_{i=1}^N$.

\subsection{FFM-TS-VAD system}

To better utilize the complementarity across acoustic features, spatial-related and speaker-related features, we propose a new FFM-TS-VAD system where a multi-level feature fusion mechanism was developed. A detailed description for FFM-TS-VAD system is displayed in Section \ref{sec:ffm}

\begin{figure}
  \centering
  \vspace{-0.5em}
  \includegraphics[width=7cm]{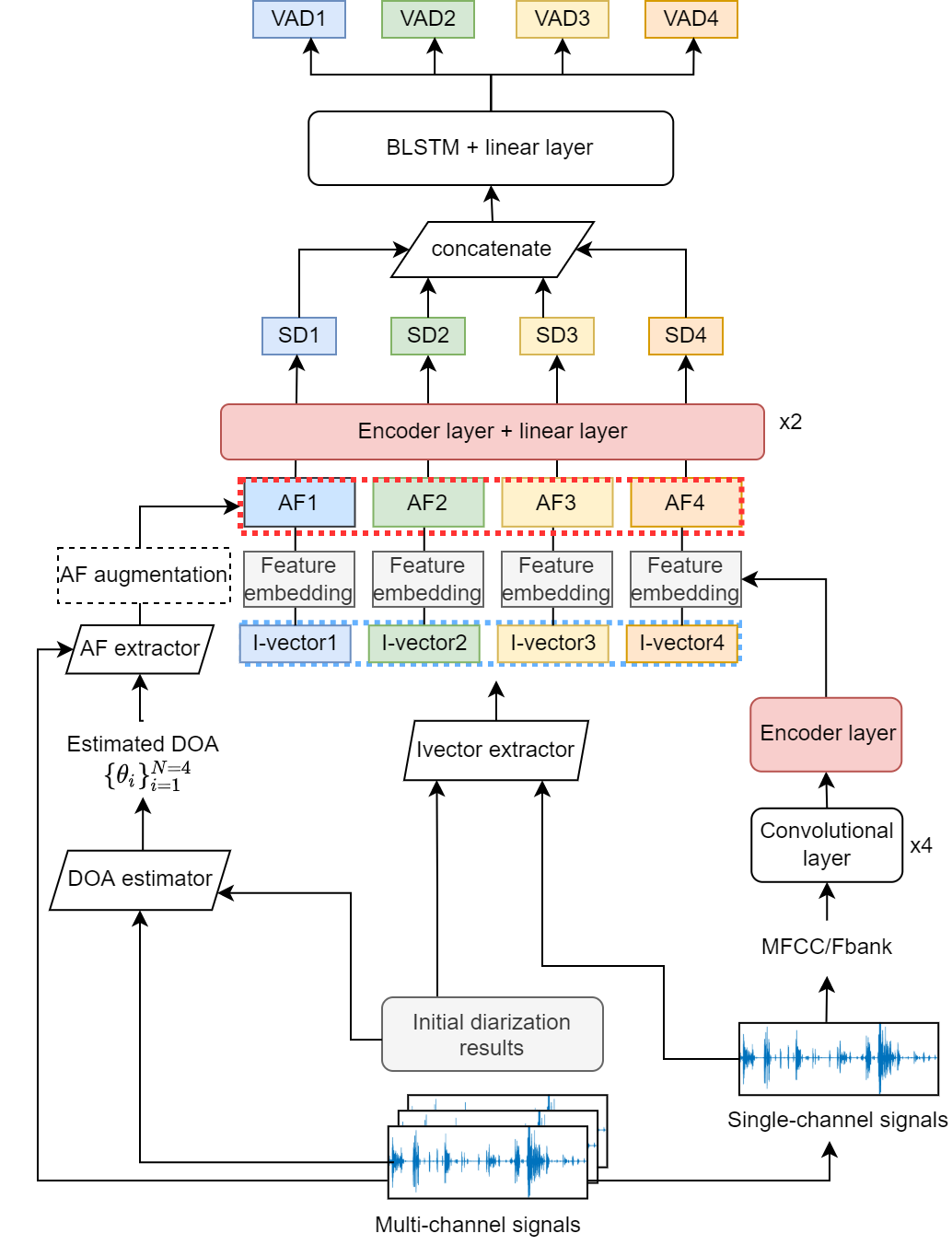}
  \caption{Diagram of the FFM-TS-VAD network.}\label{tsd_detector}
\end{figure}
\vspace{-1.0em}

\section{Proposed FFM-TS-VAD system}
\label{sec:ffm}

Figure \ref{tsd_detector} shows the diagram of the proposed FFM-TS-VAD system. Features of different levels, such as the acoustic-related feature embedding, location-guide angle feature (AF) and speaker-related i-vector, are fused as the inputs. The speaker detection (SD) information is then processed by two encoder layers and a linear layer as that in the modified TS-VAD system. Finally, the activity outputs are produced by feeding the concatenated SDs to an BLSTM layer followed by a linear layer.

%We first describe the computation of the spatial features, then we discuss its limitations when the angular difference between two speakers is relatively small.
%At last, we describe our methods to alleviate this problem.

\begin{figure}
  \centering
  \vspace{-0.5em}
  \includegraphics[width=6.5cm]{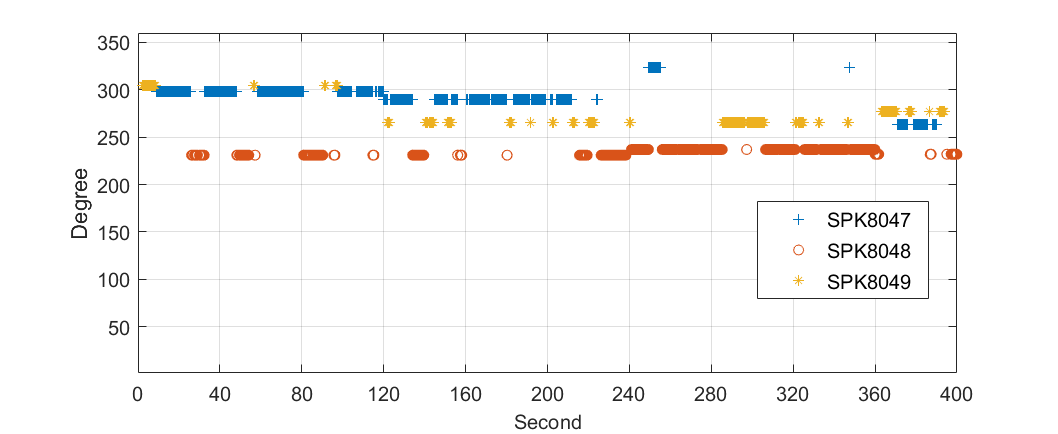}
  \vspace{-0.5em}
  \caption{The track of the DOA estimates in R8008\_M8013 meeting.}\label{DOA_conf}
\end{figure}

\subsection{Input features}
\label{ssec:doa}

The input include three types of features, i.e., the acoustic features, the AF and the i-vector. 
The following introduces the computation details of the used AF and i-vector.
% the following details the computation of AF features and i-vector.

\subsubsection{Computation of AF}
\label{sssec:af}
We first estimate the DOAs of target speakers, and then compute AF based on the estimated DOAs and multi-channel signals.
To obtain accurate DOAs of the speakers, the time distribution of speakers' activities is a prerequisite.
During the training stage, we use the reference annotations to obtain speaker activity information.
While, during the evaluation stage, we use the initial diarization annotations from our baseline system.
According to the annotations, we collect the active segments of each speaker and filter out  the overlapped speech and the short-duration segments.
Since the speakers in M2Met dataset were required to remain in the same positions, we then concatenate all the speech segments of the target speaker to estimate the DOAs.
Given the geometric information of the microphone array, FRI-based DOA estimation algorithm \cite{pan2017frida} is implemented for the estimation.
% which has high resolution at extremely low signal-to-noise ratios

Given the DOA $\theta$ and multi-channel signals ${\textbf{x}(t)}$,  %we can compute AF.
we first compute the logarithmic power spectrum (LPS) $\textbf{X}(t,f)$ from the magnitude of the spectrograms from the waveform of channels.
Then, we indicate several microphone pairs $\{\bar{m}=(m_1,m_2)\}$ and compute the inter-channel phase difference (IPD) as follows:
\begin{equation}\label{IPD}
  \text{IPD}^{\bar{m}}(t,f) = \angle{X_{m_1}(t,f)} - \angle{X_{m_2}(t,f)},
\end{equation}
where $\angle{X_{m}}$ denotes the phase of the complex spectrogram obtained from the $m$-th channel.
Finally, AF can be computed as:
\begin{equation}\label{AF}
  AF_\theta(t,f) = \sum\nolimits_{\bar{m}}\cos(\angle{v_{\theta}^{\bar{m}}(f)}-\text{IPD}^{\bar{m}}(t,f)),
\end{equation}
where $\angle{v_{\theta}^{\bar{m}}}(f) = 2\pi f\Delta^{\bar{m}}\cos(\theta^{\bar{m}})/c $ denotes the phase differences between the selected microphone pair for direction $\theta$ at frequency $f$, $\Delta^{\bar{m}}$ is the distance between the selected microphone pair, $\theta^{\bar{m}}$ is the relative angle between the look direction $\theta$ and the microphone pair $\bar{m}$, and $c$ is the sound velocity. 
% Given ${\theta_i}, i=1,...,N$ from \textit{i}-th speaker, $AF_{\theta_i}$ contains more specific spatial information about the sound intensity than the estimated DOA $\theta_i$ at time $t$. 
Given ${\theta_i}, i=1,...,N$ from \textit{i}-th speaker, $AF_{\theta_i}$ contains more specific spatial information about the sound intensity than the estimated DOA $\theta_i$ at time $t$. 
In other words, if the dominating sound is from $\theta$, then $AF_\theta$ will be close to 1.

\subsubsection{Computation of i-vectors}
The extraction of i-vectors was based on a gender-independent universal background model (UBM) with 512 mixtures and a total variability matrix with 100 total factors. 
64-dimensional FBank and 40-dimensional MFCC were used to train the UBMs and total variability matrices, respectively.

%\subsection{Confusion from close speakers' locations}
%\label{ssec:confusion}

\subsection{Data augmentation strategy for AF}
\label{augmentation}
When spatial information is used for speaker diarization, the system performance degrades a lot in presence of those cases where different speakers are very close. Figure \ref{DOA_conf} shows the tracks of the DOA along the time axis, where SPK8047 and SPK8049 had very close DOA. Since close DOA led to similar AF, it misled the network and caused it to lose the capability to discriminate the close speakers. 
%Thus, it is important to develop a method to address this problem when using the spatial features.

Considering the complementary information between the AF and i-vectors, we aim to develop a network that can adaptively select reliable features for the difficult cases above.
%However, there are few sessions whose speakers have small angular differences.
To increase the data for such cases based on the training set, it is necessary to exploit a data augmentation strategy during the training stage.
The conventional method of using near-field recordings and artificial reverberation impulse to simulate multi-channel signals have high computational costs and may incur mismatch with the real-recorded data.
% We propose an efficient approach to simulate the close location via the spatial features during training.
We propose an efficient approach to simulate spatial features of closed speakers during training.

Given a speaker and the estimated DOA $\theta_i$ in a meeting recording, we first compute the corresponding $AF_{\theta_i}$.
Imagining that when speaker $i$ and speaker $j$ get closer, their originally independent AF will become more similar. %interleave with the other and
When they get close enough, anyone's sound may lead to high AF. 
Thus, we use $\max(\cdot)$ operation to simulate this condition, where the AF features from speaker $i$ and speaker $j$ can be roughly replaced with $\hat{AF}$ as follows: 
\begin{equation}\label{af_conf}
\begin{split}
\hat{AF}_{\theta_i}(t, f) = \max(AF_{\theta_i}(t,f), \gamma \cdot AF_{\theta_j}(t,f))  \\
\hat{AF}_{\theta_j}(t, f) = \max(\gamma \cdot AF_{\theta_i}(t,f), AF_{\theta_j}(t,f)) ,
\end{split}
\end{equation}
where the parameter $\gamma$ controls the similarity degree. 
In this way, the DOAs of two speakers can also be updated as $\theta_{i,j} \leftarrow (\theta_i +\theta_j)/2 + {\theta_n}$, where $\theta_n$ is a random perturbation of small value.

To give more specific information to the network, we also compute the minimum angular difference of each speaker as follows:
\begin{equation}\label{doa_diff}
\vspace{-0.3em}
    \Delta\theta_i = \min_{k\in [1,N], k\neq i} |\theta_i - \theta_k|, 
\vspace{-0.3em}
\end{equation}
which are concatenated with the AF features as input.

%\subsection{Feature fusion}
%\label{ssec:fusion}
%We first investigate the feature fusion methods based on the spatial feature and the acoustic embedding features to alleviation the location confusion problem.
% The ideas behind these networks are similar, that is, use long-term information extracted from the whole meeting to refine the results in the overlapped speech.
%As shown in Figure \ref{tsd_detector}, the feature embeddings are extracted from the acoustic inputs with convolution layers and a transformer encode layer.
%Then, we concatenate the feature embeddings with the i-vectors and AF for each speaker and input them to the shared transformer encoder layers to obtain the speaker detection(SD) information. Finally, we combine the SD information for all speakers by a BLSTM layer and obtain their activity outputs respectively.

\subsection{Training loss function of the network}
Since the order of the outputs for speakers matches with the order of input DOAs and i-vectors, we do not need to do permutation-invariant training.
To pay more attention to the overlapped speech, we use a multi-objective loss function.
First, the activities of the speech $\hat{y}^\text{vad}$ and overlapped speech $\hat{y}^\text{osd}$ can be determined from the maximum value and the second maximum value of ${\bf{\hat{y}}^\text{vad}}=\{\hat{y}^\text{vad}_i\}_{i=1}^N$ respectively, i.e.,
\begin{equation}\label{osd}
\vspace{-0.3em}
    \hat{y}^\text{vad} = \max\{\hat{y}^\text{vad}_i\}_{i=1}^N,\ \ \hat{y}^\text{osd} = \max_{2nd}\{\hat{y}^\text{vad}_i\}_{i=1}^N
\vspace{-0.3em}
\end{equation}
Then, the training loss function can be written as
\begin{equation}\label{loss}
\begin{split}
  loss &= {\text{BCE}}({\bf{\hat{y}}^\text{vad}, \bf{y}^\text{vad}}) + \alpha * {\text{BCE}}(\hat{y}^\text{vad}, y^\text{vad}) \\
    &+ \beta * \text{BCE}(\hat{y}^\text{osd}, y^\text{osd})
\end{split}
\end{equation}
where the first component loss is the binary cross entropy (BCE) loss of activities for each speaker, and the remaining loss components expect the network to focus more on VAD and OSD with weights $\alpha$ and $\beta$.

% \subsection{Resegmentation algorithm for overlapped speech}
% \label{sec:resegmentation}
The speaker assignment algorithm can be simply achieved by a threshold-based algorithm on ${\bf{\hat{y}}^\text{vad}}$.
To obtain the final segments, the short burrs and gaps can be simply discarded and filled respectively with a post-processing step \cite{Bredin2021}.
% , where we use two thresholds for VAD and OSD respectively.
% When given the oracle VAD bounds, we only use the threshold for OSD, and select the speakers with largest probability for voiced regions.

Table generated by Excel2LaTeX from sheet 'Sheet4'
\begin{table*}[!htbp]
  \centering
  \caption{Diarization results in term of DER(\%) and JER(\%) with oracle VAD for the evaluation set and test set. SpC: Spectrum cluster. FA: false alarm. SC: speaker confusion. $(\cdot)\star$ denotes our baseline system to provide the initial diarization annotations. ($\cdot$)* denotes the meeting list which excludes the meetings whose speakers have small angular differences($\leq 45\degree$). }
  \scalebox{0.75}{
\begin{tabular}{c|c|cc|c|ccc|cc|cc|cc|cc}
\hline
\multirow{2}[4]{*}{No.} & \multirow{2}[4]{*}{Model} & \multicolumn{2}{c|}{\multirow{2}[4]{*}{Input features}} & \multirow{2}[4]{*}{AF aug.} & \multicolumn{5}{c|}{Sub-track1 on Eval.} & \multicolumn{2}{c|}{Sub-track1* on Eval.} & \multicolumn{2}{c|}{Sub-track2 on Eval.} & \multicolumn{2}{c}{Sub-track1 on Test.} \\
\cline{6-16}      &       & \multicolumn{2}{c|}{} &       & FA    & MISS  & SC    & DER   & JER   & DER   & JER   & DER   & JER   & DER   & JER \\
\hline
1     & SpC+VBx & \multicolumn{2}{c|}{X-vector} & -     & 0.00  & 13.10  & 0.47  & 13.57  & 25.10  & 14.36  & 25.94  & 13.60  & 25.31  & 13.51  & 25.73  \\
2$\star$     & SpC+VBx+OSD & \multicolumn{2}{c|}{X-vector} & -     & 1.09  & 4.17  & 2.84  & 8.10  & 20.87  & 8.25  & 21.19  & 8.00  & 20.80  & 9.45  & 21.96  \\
\hline
3     & TS-VAD & \multicolumn{2}{c|}{Fbank-based i-vector} & -     & 1.81  & 2.68  & 1.27  & 5.76  & 14.70  & 5.98  & 14.98  & 5.80  & 14.68  & 7.04  & 15.97  \\
4     & TS-VAD & \multicolumn{2}{c|}{MFCC-based i-vector} & -     & 1.74  & 2.62  & 1.10  & 5.46  & 14.41  & 5.67  & 14.78  & 5.50  & 14.32  & 6.92  & 15.92  \\
5     & Target-DOA & \multicolumn{2}{c|}{LPS+IPD+AF} & -     & 1.17  & 3.92  & 4.15  & 9.23  & 17.34  & 4.78  & 12.21  & 9.84  & 17.90  & 11.95  & 20.75  \\
\hline
6     & FFM-TS-VAD & \multicolumn{2}{c|}{AF+Fbank-based i-vector} & \XSolidBrush     & 1.04  & 2.86  & 0.34  & 4.24  & 11.55  & 3.86  & 10.82  & 4.49  & 11.79  & 7.77  & 15.45  \\
7     & FFM-TS-VAD & \multicolumn{2}{c|}{AF+MFCC-based i-vector} & \XSolidBrush     & 1.08  & 3.10  & 0.39  & 4.47  & 11.83  & 4.01  & 11.75  & 4.53  & 12.01  & 6.79  & 15.03  \\
8     & FFM-TS-VAD & \multicolumn{2}{c|}{AF+Fbank-based i-vector} & \Checkmark     & 0.83  & 2.55  & 0.26  & \textbf{3.64} & \textbf{10.53}  & \textbf{3.31} & \textbf{10.01}  & \textbf{3.91} & \textbf{10.75}  & \textbf{5.63} & \textbf{12.75}  \\
9     & FFM-TS-VAD & \multicolumn{2}{c|}{AF+MFCC-based i-vector} & \Checkmark     & 1.07  & 2.93  & 0.35  & 4.34  & 11.71  & 3.77  & 10.86  & 4.37  & 11.75  & 7.06  & 14.43  \\
\hline
      & \textit{Dover-Lap} & \multicolumn{2}{c|}{\textit{System 2~9}} & \textit{-} & \textit{0.49} & \textit{2.57} & \textit{0.22} & \textit{3.28} & \textit{10.28} & \textit{-} & \textit{-} & \textit{3.38} & \textit{10.63} & \textit{4.60} & \textit{12.39} \\
      & \textit{Dover-Lap} & \multicolumn{2}{c|}{\textit{System 2~4, 5*,6*,7*, 8~9}} & \textit{-} & \textit{0.51} & \textit{2.49} & \textit{0.23} & \textit{3.22} & \textit{10.23} & \textit{-} & \textit{-} & \textit{3.23} & \textit{10.42} & \textit{3.98} & \textit{11.19} \\
\hline
      & \textit{Official baseline []} & \multicolumn{2}{c|}{\textit{-}} & \textit{-} & \textit{-} & \textit{-} & \textit{-} & \textit{15.24} & \textit{-} & \textit{-} & \textit{-} & \textit{-} & \textit{-} & \textit{15.67} & \textit{-} \\
\hline
\hline
    \end{tabular}%
    }
  \label{tab:der}%
  \vspace{-1.5em}
\end{table*}%

\section{Experimental SETUP}
\label{sec:experiments}

\subsection{Data preparation}
\label{ssec:data_pre}
For sub-track 1, we used the AliMeeting, AISHELL-4 and CN-Celeb to train our speaker embedding extractors, as a result the training set contains 3,544 speakers in total.
%For the AliMeeting corpus and the AISHELL-4 corpus, we collect the speakers' utterances based on their reference annotations.
During training OSD and network-based diarization systems, only the AliMeeting and the AISHELL-4 corpora were used.

For sub-track 2, besides the above three corpora, the Voxceleb 1\&2 \cite{nagrani2017voxceleb, chung2018voxceleb2} corpora were also used to train the speaker embedding extractor in our baseline system.
The other systems remained the same with sub-track 1.

\subsection{Data augmentation for training set}

During training the speaker embedding extractor, i-vector extractors and TS-VAD networks, 3-fold speed perturbation (x0.9, x1.0, x1.1) was applied to increase the number of speakers.
We only performed data augmentation with the noise part in the MUSAN corpus and no reverberation was used.

While training the OSD and network-based diarization systems, we applied data augmentation to balance the probability of overlapped speech and non-overlapped speech using on-the-fly mode.
50\% of the training segments were artificially made by summing two chunks cropped from the same meetings, and the signal-to-signal ratios were sampled between 0 and 10dB.
During training the FFM-TS-VAD systems, 50\% of the training segments were artificially augmented with Eq \eqref{af_conf} when using the AF augmentation.

% However, even with the reference segmentation, the estimation method can give no guarantee that all DOA estimation are correct. 
% If use wrong DOA information as input during training, these ``dirty data" will deeply  damage our system.
% Thus, we conduct an additional process to discard these unreliable estimates, where we divide the 8-channel microphone array as two 4-channel microphone sub-arrays by interval selecting microphones.
% Then, we compare the estimated DOA from the original 8-channel array and the two sub-arrays. 
% Once they have large mismatch, we will remove the meeting from training dataset.

\subsection{Implementation details}
In our baseline systems, the threshold in the re-clustering process was set to 0.7.
When computing the IPD and AF, we selected 4 microphone pairs, i.e., (1, 5), (2, 6), (3, 7) and (4, 8) in Eq \eqref{IPD} and Eq \eqref{AF}.
All the networks were trained on short segments with a fixed 4-second length, and the number of the output ports $N$ was set to 4.
During evaluation, if the actual number of speakers was less than 4, we selected the speaker embeddings from the independent training dataset as `virtual' speakers and selected directions with large angular differences from the target speakers as `virtual' DOAs. 

We used the transformer layers with 4 multi-heads and 512 units in the modified TS-VAD and FFM-TS-VAD systems while keeping the other configurations same to \cite{Medennikov2020}.
The $\gamma$ in Eq \eqref{af_conf} was uniformly sampled within $[0.8, 1]$ and $\theta_n$ was randomly sampled from $[0, 45\degree]$.
For the loss function in Eq \eqref{osd}, $\alpha$ and $\beta$  were manually set to 0.25.
For all diarization networks, the learning rate was set to 1e-4 with Adam optimizer.
ReduceLROnPlateau schedule and early stop were also adopted.

% To obtain stable DOA estimates for target-DOA systems, the duration threshold to filter out short active segments is empirical set to 5 seconds.

\section{Results and Analyses}
\label{sec:results}

% \subsection{Speaker diarization with handling overlaps}

Table \ref{tab:der} shows the diarization results for the evaluation set in terms of diarization error rate (DER) and Jaccard error rate (JER), where the collar is set to 0.25 seconds.
Oracle VAD is used in all systems and all sub-tracks.

We applied beamforming for System 1$\sim$4 with the BeamformIt tool \cite{anguera2007acoustic} to obtain enhanced single-channel speech.
Since System 1 does not handle the overlaps, the DER is mainly from the miss rate.
With the OSD applied in System 2, the miss rate decreases a lot, but the speaker confusion (SC) error increases due to inaccurate overlapped speaker identification based on the heuristic algorithm.
Compared with above systems, TS-VAD systems (System 3 and 4) can obtain much better performance, especially in SC. 
% Different with \cite{Medennikov2020} which re-estimates i-vectors iteratively, we only use single iteration to extract the i-vectors and DOAs in System 3$\sim$9, since the initial diarization annotations obtained from System 2 already indicates the overlapped speech regions.
We used single iteration to extract the i-vectors and DOAs in System 3$\sim$9, since the initial diarization annotations obtained from System 2 already indicate the regions of overlapped speech.

System 5 uses the spatial information to handle overlapped speech.
As discussed earlier, when the angular difference between the speakers is relatively small, the computed AF cannot identify the speakers correctly.
To avoid such bad cases, we exclude the meetings\footnote{R8008\_M8013 is excluded, and 7 other meetings are remained.} whose speakers have small angular differences ($\leq 45\degree$), which can be detected according to Eq \eqref{doa_diff}.
The results on the remaining meetings are listed in the sub-track1* column in Table \ref{tab:der}.
In this case, the DER of System 3 is better than those of the TS-VAD systems, which shows the high efficiency of spatial information in most situations.

For System 6$\sim$9, we applied the proposed FFM-TS-VAD system using both AF and i-vectors.
With feature fusion, the networks can obtain further reduction on the SC error and obtain much more accurate estimation of overlaps compared to TS-VAD and Target-DOA systems.
Meanwhile, when AF augmentation strategy is applied, System 8 with AF and FBank-based i-vectors inputs can obtain relative 55.06\% improvement of DER against System 2.
% and 9 can give reliable estimation on the evaluation set. % based on the feature quality.
% Finally, System 8 with AF and FBank-based i-vectors inputs can obtain relative 55.06\% improvement of DER against System 2.

We also list the results in sub-track2 in Table \ref{tab:der}. Compared with the systems in sub-track1, we only use a different speaker embedding extractor in System 1 and 2 while keeping the other networks unchanged, and similar results can be obtained.

Finally, we applied the Dover-lap algorithm with greedy mapping \cite{raj2021dover} to do system fusion and the absolute DER on evaluation set can reach 3.28\%.
When we use the results from sub-track1* for System 5$\sim$6 during fusion, we observe a further reduction on DER which can be reduced to 3.22\% finally.
For the test set, the DER of our fusion system achieves 3.98\%. Compared with the official baseline system \cite{yu2021m2met} whose DER is 15.24\% on evaluation set and 15.6\% on test set, our fusion system achieves significant improvement.

\section{CONCLUSION AND FUTURE WORK}
\label{sec:con}
In this paper, we present our system for overlap-aware diarization tasks in the M2Met challenge.
% Several speaker assignment algorithms are investigated for overlapped speech.
The proposed FFM-TS-VAD system using the multi-level feature fusion mechanism and the data augmentation strategy obtains marked improvement over conventional methods.
Processing confusable spatial features extracted from the close speakers in the multi-channel speaker diarization task remains an open problem.
% It is still an open problem to process the ambiguous spatial features extracted from the close speakers in multi-channel speaker diarization task.
% In the future, we will attempt to use simulated multi-channel speech for training.

\vfill\pagebreak

\bibliographystyle{IEEEbib}
\bibliography{strings,refs}

\end{document}